\documentclass[final,3p,times,twocolumn]{elsarticle}
\usepackage{bm}
\usepackage{placeins}
\usepackage{amsfonts}
\usepackage{multirow}
\usepackage{graphicx,color,amsmath,amssymb}
\usepackage[utf8]{inputenc}

\newcommand{\eg}{\textit{e.g.}}

\newcommand{\ie}{\textit{i.e.}}

\newcommand{\lsim}{\lesssim}

\biboptions{sort&compress}

\begin{document}

\title{Extraction of the Heavy-Quark Potential from Bottomonium Observables in Heavy-Ion Collisions}

\author{Xiaojian Du, Shuai Y.F. Liu and Ralf Rapp}
\address{Cyclotron Institute and Department of Physics and Astronomy, 
	Texas A$\&$M University, College Station, Texas 77843-3366, USA
}

\date{\today}

\begin{abstract}
The in-medium color potential is a fundamental quantity for understanding the properties of the strongly 
coupled quark-gluon plasma (sQGP). Open and hidden heavy-flavor (HF) production in ultrarelativistic 
heavy-ion collisions (URHICs) has been found to be a sensitive probe of this potential.
Here we utilize a previously developed quarkonium 
transport approach in combination with insights from open HF diffusion to extract the color-singlet potential 
from experimental results on $\Upsilon$ production in URHICs. Starting from a parameterized trial potential,
we evaluate the $\Upsilon$ transport parameters and conduct systematic fits to available data for the 
centrality dependence of ground and excited states at RHIC and the LHC. The best fits and their statistical
significance are converted into a temperature dependent potential. Including nonperturbative effects in the
dissociation rate guided from open HF phenomenology, we extract a rather strongly coupled potential with 
substantial remnants of the long-range confining force in the QGP.
\end{abstract}


\maketitle

\section{Introduction} 
The confining force of Quantum Chromodynamics (QCD) plays a central role in the 
quantitative description of the bound-state spectra of charmonia and bottomonia in 
vacuum~\cite{Eichten:1974af}, characterized by a linear term in color-singlet potential
between a color charge and its anti-charge. It has also been applied rather successfully
for light hadrons~\cite{Godfrey:1985xj,Lucha:1991vn} (with caveats in the chiral sector).
Thus, the in-medium properties of quarkonia have long been recognized as promising probe for 
the formation of the quark-gluon plasma (QGP) in ultrarelativistic heavy-ion collisions 
(URHICs)~\cite{Matsui:1986dk,Kluberg:2009wc,BraunMunzinger:2009ih,Rapp:2008tf,Mocsy:2013syh,Rapp:2017chc}.
In addition, the consequences of in-medium potentials on heavy-quark (HQ) 
diffusion~\cite{vanHees:2007me,Riek:2010fk,Liu:2018syc} and QGP 
structure~\cite{Brown:2003km,Shuryak:2004tx,Mannarelli:2005pz,Liao:2008vj,Liu:2016ysz,Liu:2017qah} 
have been studied, where remnants of the confining force above the pseudo-critical temperature, 
$T_{\rm pc}$, were found to be essential in explaining the properties of the strongly-coupled QGP 
(sQGP). 
Recent efforts to define the potential~\cite{Laine:2006ns,Beraudo:2007ky,Brambilla:2008cx} 
and relate it to quantities computed in lQCD, such as the free energy or quarkonium correlators, 
made progress in extracting this potential~\cite{Rothkopf:2011db,Burnier:2014ssa,Liu:2015ypa}. 
In these approaches the HQ free ($F$) and internal ($U$) energies, previously used as potential 
proxies, are rather outputs of suitably defined interaction kernels. However, the present results 
are not unique, ranging from a weak potential~\cite{Burnier:2014ssa,Liu:2017qah}, close 
to $F$, to a stronger one~\cite{Bazavov:2014kva,Liu:2015ypa,Liu:2017qah}, close 
to the vacuum potential at moderate QGP temperatures. 

Transport analyses of open and hidden heavy-flavor (HF) production in URHICs 
require relatively strong interaction potentials for heavy-light and heavy-heavy interactions, 
respectively. For example, low-momentum $D$-meson observables (in particular their elliptic
flow) clearly favor the $U$-potential over the $F$-potential proxy~\cite{Riek:2010fk,Liu:2018syc}
(or require large $K$ factors when using perturbative interactions~\cite{Rapp:2018qla}); similar
trends are found for quarkonium observables in 
URHICs~\cite{Zhao:2010nk,Liu:2010ej,Emerick:2011xu,Strickland:2011aa,Zhou:2014hwa} 
albeit systematic constraints have not been evaluated yet. 
Bottomonium observables are particularly promising to achieve that. 
Theoretically, the large bottom-quark mass renders the potential approach 
most suitable; phenomenologically, regeneration contributions are expected to be 
smaller~\cite{Emerick:2011xu} than in the charmonium sector where they reduce the sensitivity 
to the underlying potential~\cite{Zhao:2010nk};
experimentally, the recent increase in available data and their precision, 
encompassing both ground ($\Upsilon(1S)$) and excited states ($\Upsilon(2S)$, $\Upsilon(3S)$) 
at RHIC~\cite{Adamczyk:2013poh,Adare:2014hje} and the 
LHC~\cite{Chatrchyan:2012lxa,Khachatryan:2016xxp,Sirunyan:2017lzi,Sirunyan:2018nsz,Acharya:2018mni}
has reached a point where a quantitative sensitivity to the in-medium potential
seems possible~\cite{Du:2017qkv}.

In the present work, we conduct a statistical analysis of the centrality dependence of 
available bottomonium data at RHIC and the LHC, with the goal of constraining the HQ 
potential at finite temperature. Toward this end, we employ our previously developed 
semi-classical Boltzmann/rate equation approach which has been extensively tested by a 
wide variety of quarkonium observables from SPS via RHIC to LHC energies for both 
charmonia and bottomonia~\cite{Grandchamp:2003uw,Zhao:2010nk,Zhao:2011cv,Du:2017qkv}. 
Its results are largely consistent with other semi-classical 
approaches~\cite{Liu:2010ej,Strickland:2011aa,Ferreiro:2012rq,Zhou:2014kka,Hoelck:2016tqf,Krouppa:2017jlg,Aronson:2017ymv,Ferreiro:2018wbd},
although quantitative cross comparisons under controlled conditions remain to be 
carried out~\cite{Rapp:2017chc}.  
Furthermore, the effects of explicit quantum evolution equations for quarkonia 
are receiving increased 
attention~\cite{Akamatsu:2014qsa,Katz:2015qja,Brambilla:2016wgg,Blaizot:2018oev,Yao:2018nmy}. 
However, it has not yet been scrutinized in how far quantum effects affect the extraction 
of transport parameters, and most of the pertinent calculations do not yet employ realistic 
potentials including the string term, which plays a critical role in the dissociation processes 
even for bottomonia. The implications for the systematic uncertainty of semi-classical approaches 
will have to be elaborated in future work.

\begin{figure}[!t]
\begin{center}
\includegraphics[width=0.46\textwidth]{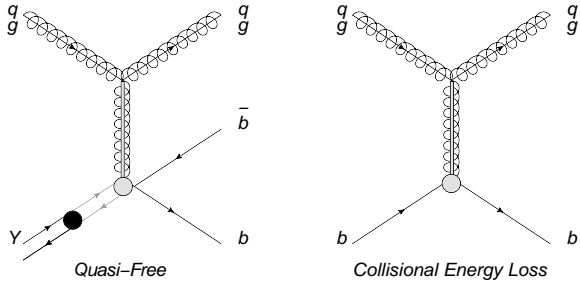}
\end{center}
\caption{Relation between ``quasifree" dissociation of quarkonia and single HQ interactions in 
the QGP. The vertical lines represent the in-medium potential between
thermal partons and heavy quark(onium).}
\label{fig_qf}
\end{figure}
The key connection between the in-medium potential and quarkonium transport is the 
inelastic reaction rate which increases as the potential weakens. In practice, inelastic 
parton ($i=q,\bar q, g$) scattering of the type $i+Y\to i+b+\bar b$ has been identified 
as the leading contribution to the dissociation rate in the relevant regime of temperatures 
where the dissociation energies
are relatively small (also referred to as an imaginary part of 
the HQ potential~\cite{Laine:2006ns,Beraudo:2007ky,Brambilla:2008cx}, or ``quasifree 
dissociation"~\cite{Grandchamp:2001pf}). Since the basic diagrams essentially
correspond to heavy-light scattering, $i+b\to i+b$, they are closely 
related to HQ diffusion, cf.~Fig.~\ref{fig_qf}. From HF phenomenology it is now well 
established that HQ transport coefficients require a large enhancement over perturbative 
results~\cite{Rapp:2018qla,Liu:2018syc}. Reliable extractions of the in-medium 
HQ potential in quarkonium transport have to account for this.

\section{$Y$ Transport and in-Medium Potential}
\label{sec_trans} 
The quarkonium transport framework employed in this work utilizes 
a rate equation~\cite{Grandchamp:2003uw,Zhao:2010nk,Du:2017qkv},
\begin{equation}
\frac{dN_Y(\tau)}{d\tau}=-\Gamma(T(\tau))\left[N_Y(\tau)-N_Y^{\rm eq}(T(\tau))\right],
\label{eq_rateeq}
\end{equation}
for the number, $N_Y$, of different bottomonia, $Y$=$\Upsilon(1S)$, $\Upsilon(2S)$, 
$\Upsilon(3S)$, $\chi_b(1P)$. The equilibrium limit, $N^{\rm eq}(T)$, governs 
regeneration processes and is obtained from the thermal model with experimental 
input for open-bottom cross sections (we also include a relaxation time correction 
for incomplete $b$-quark thermalization and correlation volume effects in the canonical 
ensemble). 
However, the regeneration contribution to bottomonia is relatively small, and $N^{\rm eq}(T)$
depends only weakly on the potential through the $b$ and $Y$ masses.

\begin{figure}[!t]
\begin{minipage}[b]{0.95\linewidth}
\hspace*{-0.3cm}
\centering
\includegraphics[width=1.12\textwidth]{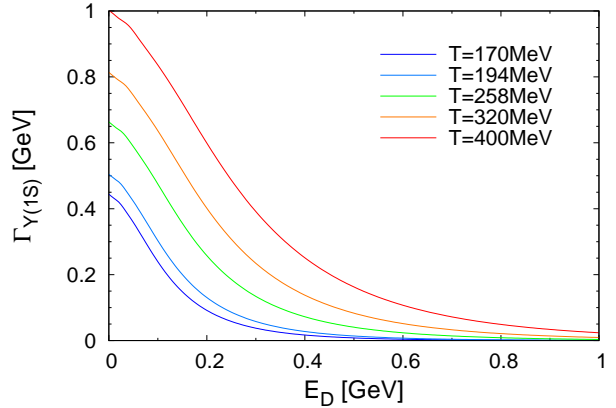}
\end{minipage}
\caption{Dissociation energy dependence of the quasifree width (with $K$=5) 
for a fixed $Y$ mass of 9.46\,GeV at different temperatures.}
\label{fig_G-E}
\end{figure}
The central quantity is the inelastic reaction rate, $\Gamma(T)$, which depends on temperature 
through the thermal-parton density and the $Y$ dissociation energy, $E_D$, which controls 
the final-state phase space of the dissociation process, cf.~Fig.~\ref{fig_G-E}.\footnote{Note
that, especially in the presence of a long-range string interaction, $E_D$ is not 
necessarily identical to the binding energy; for example, for the $Y$ ground state in vacuum, 
its binding may not be strongly affected by the string interaction, but $E_D$, determined
by the $B\bar B$-meson threshold, directly depends on it. In the present paper, we therefore 
use the notion of dissociation energy as the relevant quantity for calculating reaction rates.} 
The main 
contribution to the rate stems from quasifree dissociation~\cite{Combridge:1978kx,Grandchamp:2001pf}  
for which we include interference effects causing a $r$-dependent reduction of the widths 
increasing with the dissociation energy of the bound state~\cite{Du:2017qkv}. The much smaller 
contributions to the rate from  gluo-dissociation~\cite{Peskin:1979va,Bhanot:1979vb} are 
also accounted for.
A significant extension over our previous work~\cite{Du:2017qkv} is the implementation of
constraints from open HF phenomenology, which require HQ scattering rates in the QGP well
beyond perturbative estimates. This is done by introducing a $K$ factor in 
the quasifree reaction rate, which for simplicity we assume to be temperature- and 
momentum-averaged.  
In the presence of large dissociation widths, the issue of the onset temperature for regeneration 
reactions needs to be revisited, \ie, at what temperature in the cooling of the fireball 
bound-state formation commences. In our previous work, where the quasifree rates were relatively 
small, the default assumption was to use the vanishing of the dissociation energy, $E_D^Y(T_{\rm reg})$=0, 
to define the temperature, $T_{\rm reg}$, below which regeneration sets in. However, for dissociation 
energies much smaller than the width (for a large $K$ factor), the formation
time of the bound state becomes longer than its lifetime. Therefore, we amend the criterion for
$T_{\rm reg}$ by defining it as the temperature where the dissociation energy becomes comparable to 
the reaction rate, $E_D^Y(T_{\rm reg})$=$\Gamma_Y(T_{\rm reg})$ (as it turns out, both
criteria lead to virtually identical results for the extracted potentials, with some difference
in the composition of primordial and regeneration components for excited states). 
Above $T_{\rm reg}$ the dissociation of would-be quarkonia (\ie, primordially produced $b\bar b$ 
quarks that in a $pp$ collision would evolve into a quarkonium bound state) is still operative 
at a rate of twice the collision rate of a single $b$ quark. 
A more rigorous treatment of these issues requires a quantum evolution approach which we defer 
to future work.

The key quantity to calculate the in-medium dissociation energies is the in-medium potential $V(r,T)$
for which we adopt a screened Cornell-type potential.
For an efficient use in the 
statistical analysis discussed below, we utilize a 2-parameter ansatz for the $T$-dependence of 
the potential (akin to that in Ref.~\cite{Mocsy:2007jz}), with a Debye screened color-Coulomb term 
and a confining term whose screening is controlled by a string breaking distance, $R_{\rm SB}$,
\begin{eqnarray}
V_{Q\bar{Q}}(r)=
\begin{cases}
-\frac{4}{3}\alpha_s \  {\rm e}^{-m_Dr}/r+\sigma r \hspace{0.6cm}, r<R_{\rm SB}
\cr
-\frac{4}{3}\alpha_s \ {\rm e}^{-m_Dr}/r+\sigma R_{\rm SB} \hspace{0.2cm}, r>R_{\rm SB} \ .
\end{cases}
\label{eq_v}
\end{eqnarray}
Here, $m_D$ and $m_S\equiv1/R_{\rm SB}$ are the pertinent screening 
masses. We have checked that the sharp-cutoff version of the string term closely 
resembles the results for dissociation energies from more elaborate smooth versions as used, \eg, in 
Refs.~\cite{Riek:2010fk,Liu:2017qah}. Its advantages are an analytical evaluation of its 
partial-wave expansion (which can be done analytically) and the dependence on a single parameters
(whose temperature dependence, however, turns out to be more involved). For a given potential 
the dissociation energies are obtained from a $T$-matrix equation and subsequently serve as input 
into the reaction rate. In the spirit of the semi-classical Boltzmann approach, they are computed 
in the narrow-width approximation, while the width effects (including interference) are represented 
by the reaction rates. 

\section{Statistical Approach}
\label{sec_stat}
\begin{figure*}[!t]
	\begin{tabular}{c}
		\begin{minipage}[b]{0.32\linewidth}
			\includegraphics[width=1.12\textwidth]{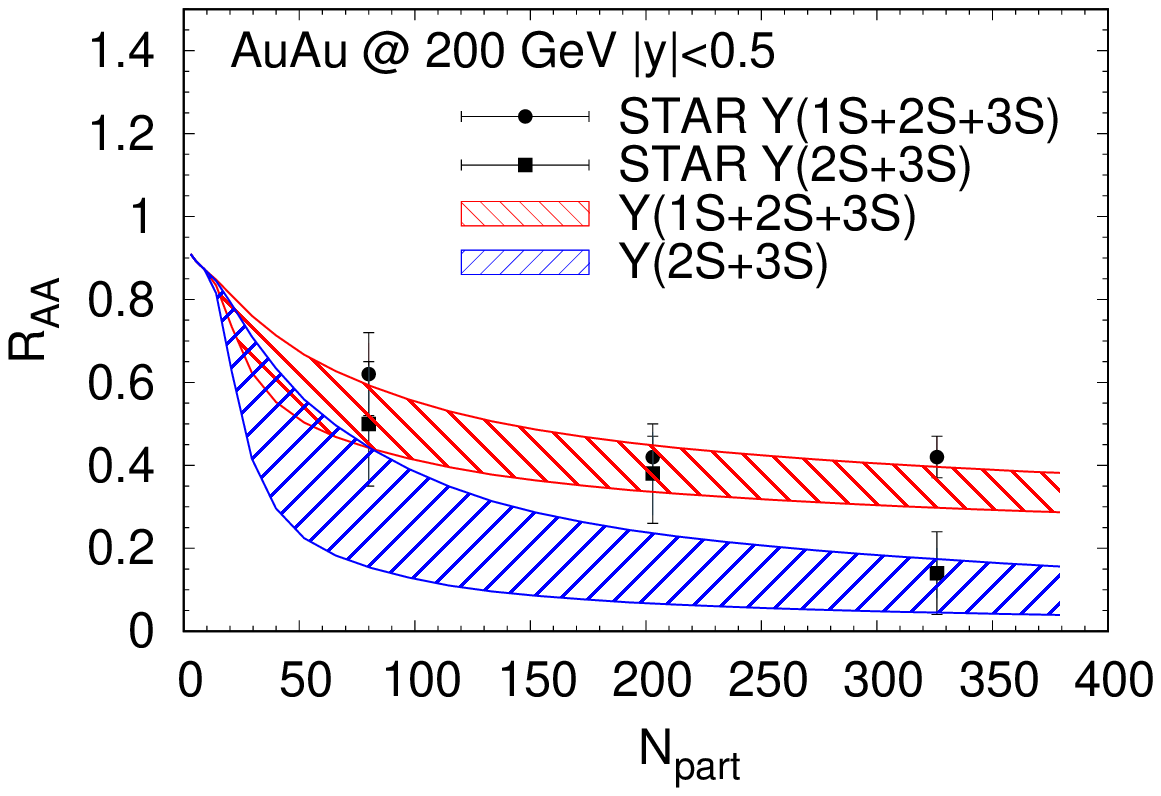}
		\end{minipage}
		\begin{minipage}[b]{0.32\linewidth}
			\includegraphics[width=1.12\textwidth]{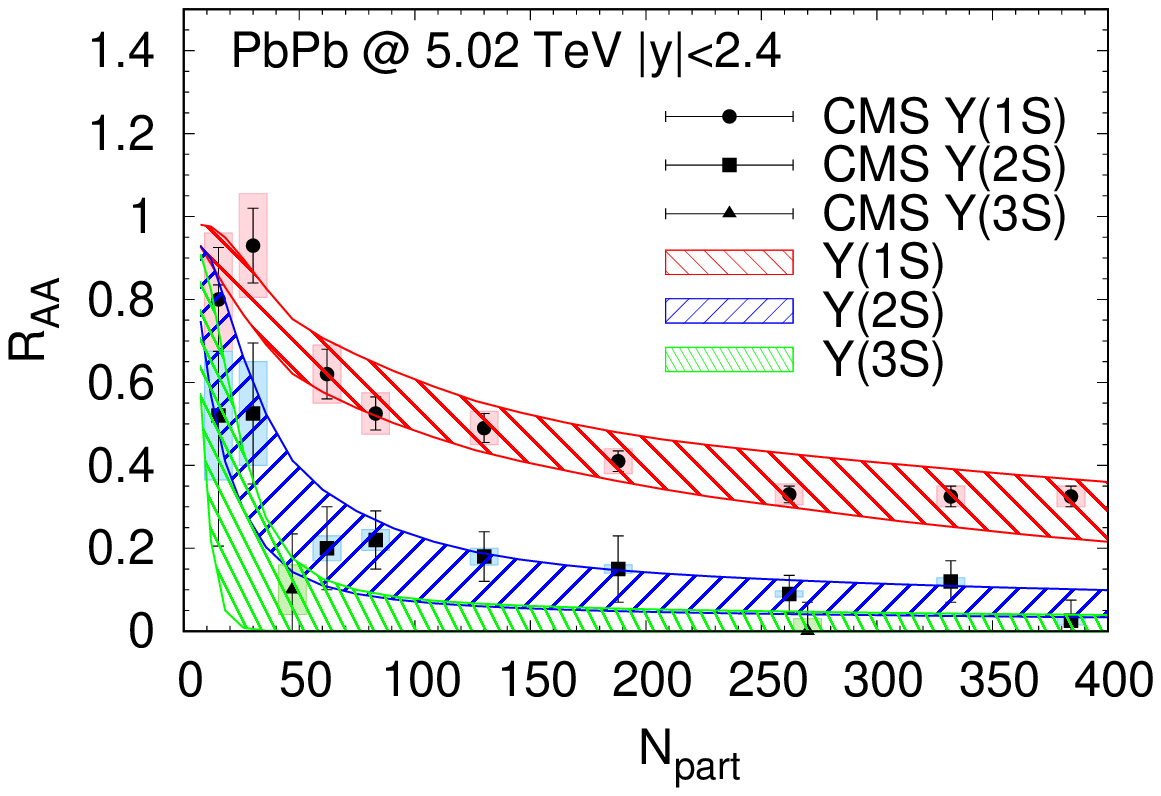}
		\end{minipage}
		\begin{minipage}[b]{0.32\linewidth}
			\includegraphics[width=1.12\textwidth]{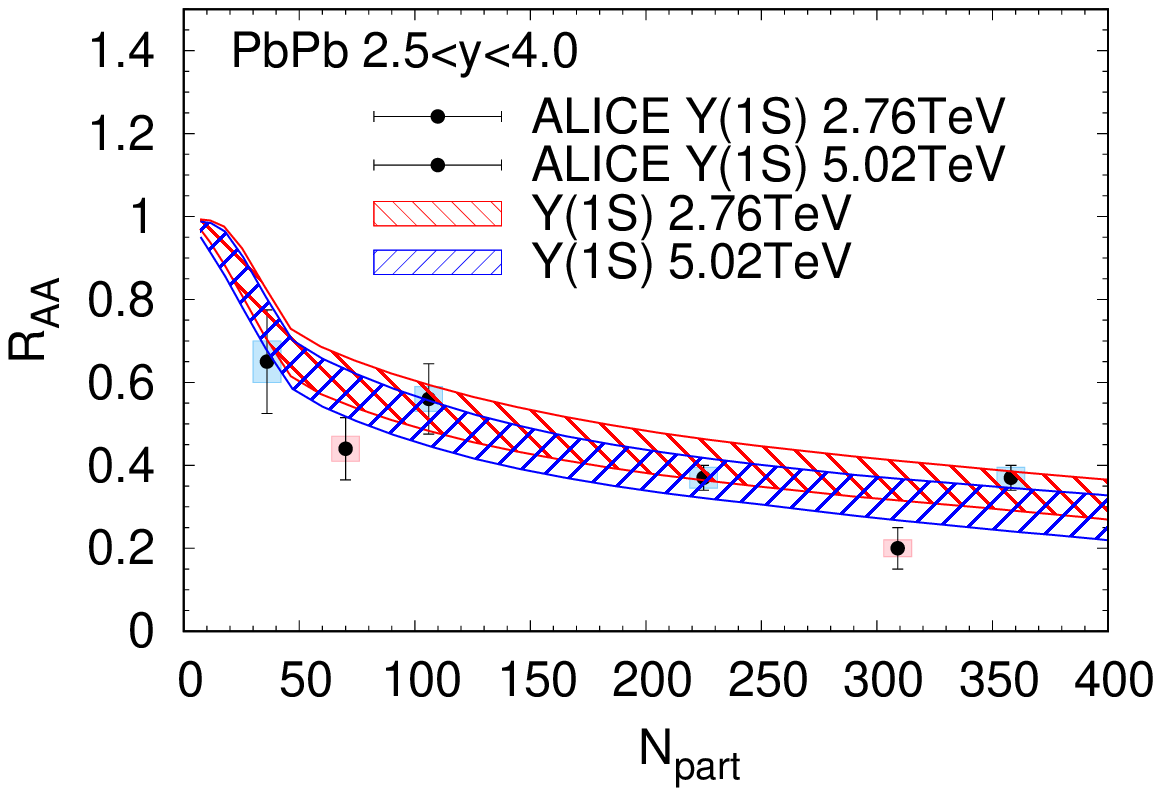}
		\end{minipage}\\
		\begin{minipage}[b]{0.32\linewidth}
			\includegraphics[width=1.12\textwidth]{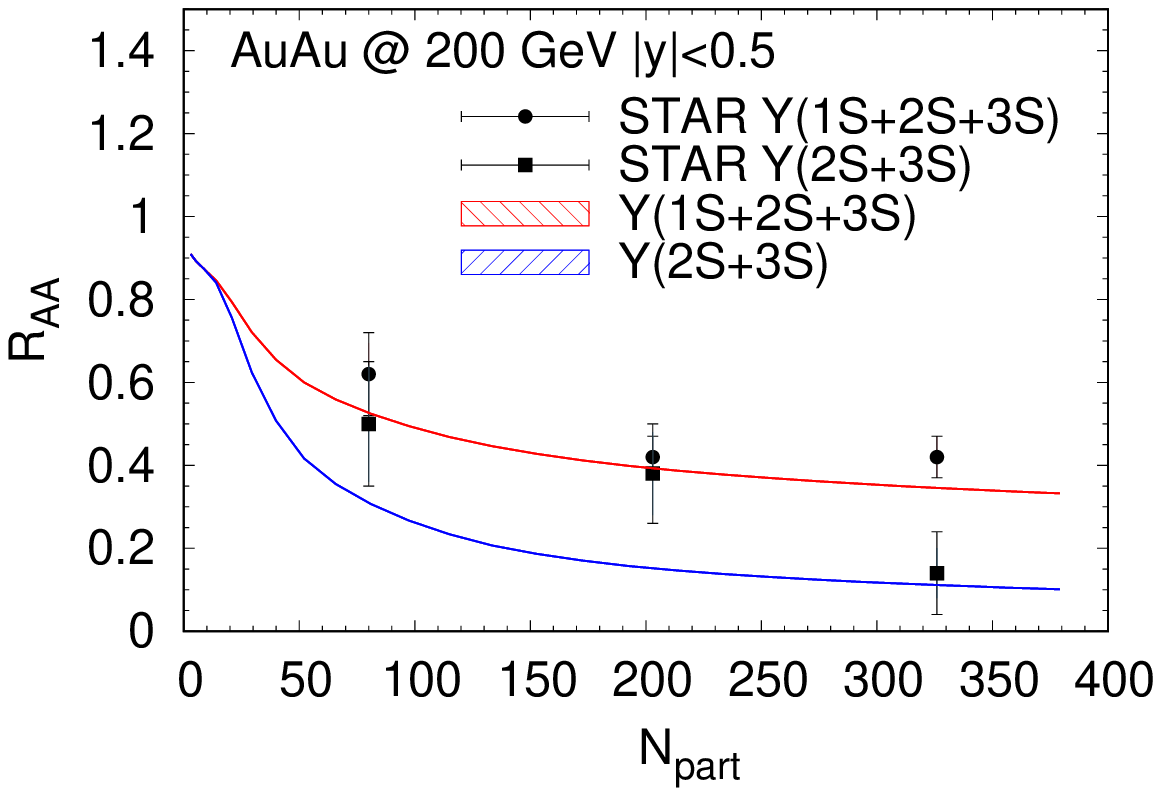}
		\end{minipage}
		\begin{minipage}[b]{0.32\linewidth}
			\includegraphics[width=1.12\textwidth]{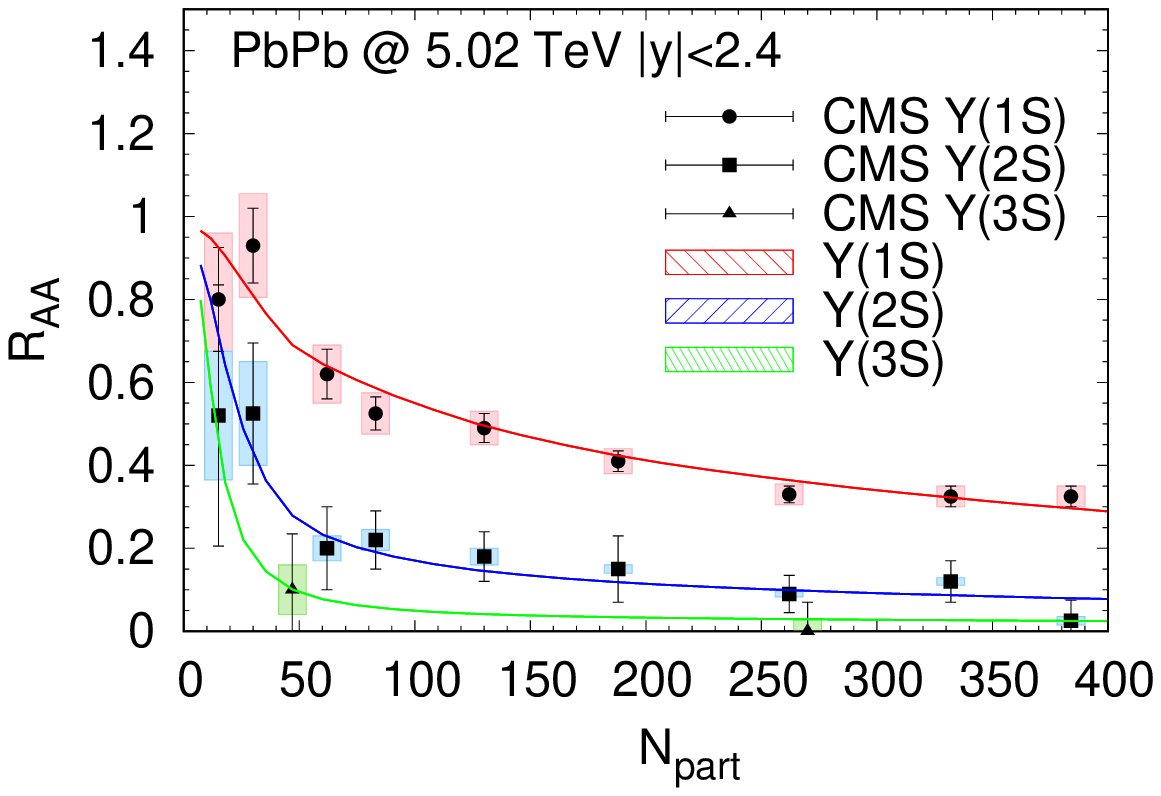}
		\end{minipage}
		\begin{minipage}[b]{0.32\linewidth}
			\includegraphics[width=1.12\textwidth]{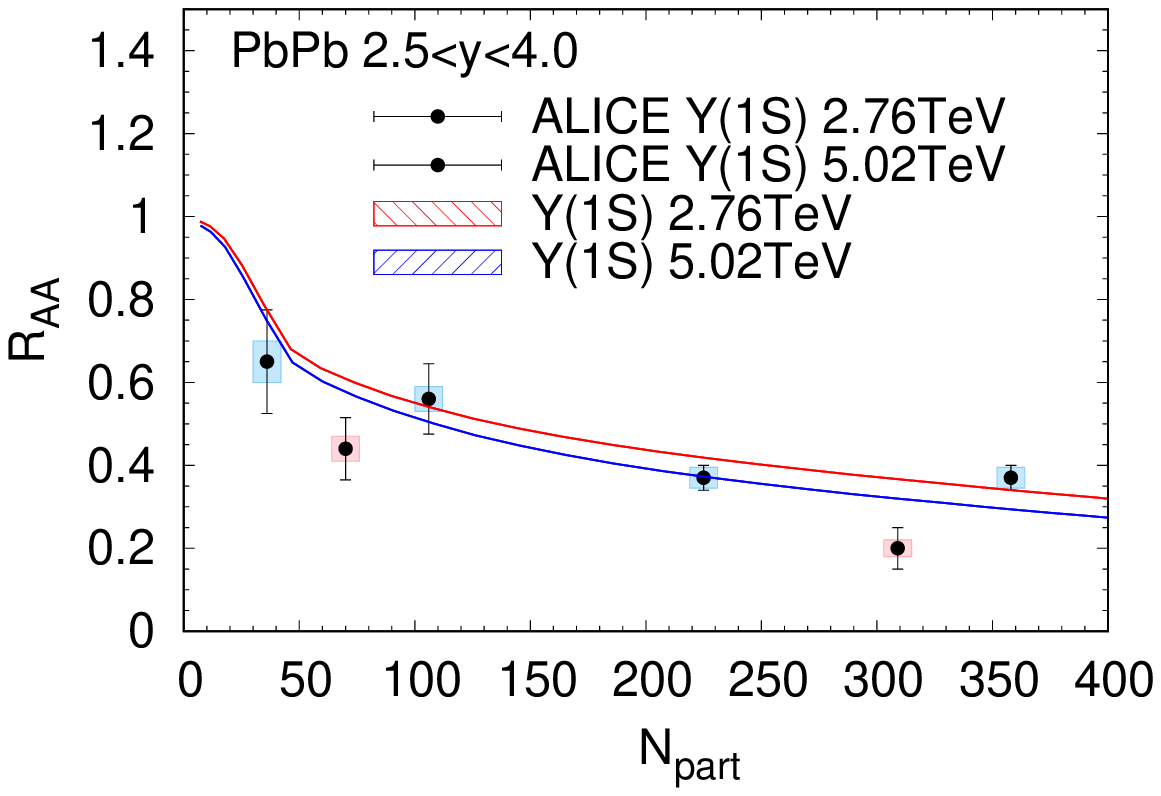}
		\end{minipage}
	\end{tabular}
	\vspace*{-0.6cm}
	\caption{Bands of 95\% confidence level (upper panels) and best-fit results (lower panels) for the $Y$ $R_{\rm AA}$'s in 
		the $K$=5 scenario, 
		compared to:  $\Upsilon(1S+2S+3S)$ and $\Upsilon(2S+3S)$ STAR data in Au-Au(0.2\,TeV) collisions (upper panels), 
		$\Upsilon(1S,2S,3S)$ CMS data at mid-rapidity in PbPb(5.02\,TeV) collisions (middle panels), and $\Upsilon(1S)$ ALICE data at
		forward rapidity in Pb-Pb(2.76,5.02\,TeV) collisions (lower panels).}
	\label{fig_raa}
\end{figure*}
To implement the in-medium potential into a statistical analysis of bottomonium data within
our transport framework, we parameterize the temperature dependence of the 
screening masses. Guided by previous studies of the potential model within the $T$-matrix 
approach, we utilize a constant strong-coupling constant, $\alpha_s$, and string tension, 
$\sigma$, together with a Debye mass linear in temperature,  
while the screening of the string term requires more flexibility. We make the ans\"atze     
\begin{align}
&m_D=aT_o\tilde{T},
\label{eq_md}
\\
&m_S=m_{S}^{\rm vac}+T_o\left[c\tilde{T}-\left(c-b\right)\left(\sqrt{\tilde{T}^2+d^2}-d\right)\right] ,
\label{eq_ms}
\end{align}
where $m_S^{\rm vac}$$\simeq$1/fm is the inverse string-breaking distance in vacuum and 
$\tilde{T}=\frac{T}{T_o}-1$ is the ``reduced" temperature relative to the onset temperature of  
screening. The four dimensionless fit  parameters characterize the  slope of $m_D$ ($a$), 
the high-$T$ and low-$T$ slopes of $m_S$ ($b$ and $c$, respectively), 
and the transition between the two ($d$); \eg, for $d$=0, the low-$T$ slope drops out. 
The in-medium $b$-quark mass includes a self energy from the potential~\cite{Riek:2010fk,Liu:2017qah},  
$m_b=m_{b}^0+\frac{1}{2}\left(-\frac{4}{3}\alpha_sm_D+\frac{\sigma}{m_S}\right)$, where 
$m_{b}^0$ is the bare mass. 
With $m_{b}^0$=4.719~GeV, $\alpha_s$=0.298, $\sigma$=0.220\,GeV$^2$ and $m_S^{\rm vac}$=0.194\,GeV 
a good fit to the vacuum masses of $\Upsilon(1S)$,  $\Upsilon(2S)$, $\chi_b(1P)$ and 
$\chi_b(2P)$ is obtained. The values for $\alpha_s$ and $\sigma$ obtained through our fit are 
consistent with lQCD results~\cite{Petreczky:2004pz,Kaczmarek:2007pb,Bazavov:2012fk} 
at the scales relevant for bottomonia.
For the onset temperature of screening, our default value is $T_{o}$=0.15\,GeV, slightly below the 
QCD pseudo-critical temperature. Guided by lQCD data~\cite{Bazavov:2013yv} for the infinite-distance 
limit of the HQ free energy below  $T_{\rm pc}$, we have also checked a smaller value of 
$T_{o}$=0.13\,GeV, but did not find significant differences in the final results for the extracted 
in-medium potential (as we will see below, the screening of the string term turns out to be small 
up to $T$$\simeq$0.2\,GeV).

For a given set of  parameters, ($a,b,c,d$), the dissociation energies of the different bottomonium 
states are calculated as a function of temperature, $E_D^Y(T;a,b,c,d)$, and the corresponding $Y$ 
masses (figuring in the equilibrium limit, $N_Y^{\rm eq}$) follow as
\begin{equation}
m_Y(T)=2 m_b(T)-E_D^Y(T) \ .
\end{equation}
With those inputs, we generate the reaction rates and evolve the $Y$ numbers through the rate equation to compute a full
set of $Y$ nuclear modification factors, 
\begin{equation}
R_{\rm AA}^Y= \frac {N^Y_{\rm AA}(N_{\rm part})} {N_{\rm coll}(N_{\rm part}) N^Y_{pp}} \ , 
\label{eq_raa}
\end{equation}
as a function of centrality (characterized by the number of nucleon participants, $N_{\rm part}$)  
at RHIC ($\sqrt{s}$=0.193,\,0.2\,TeV) and the LHC ($\sqrt{s}$=2.76,\,5.02\,TeV, at both forward and 
mid-rapidity); $N^Y_{\rm AA}(N_{\rm part})$ denotes the final $Y$ yield in an AA collision, which is
normalized to its binary-collision number-scaled yield in $pp$ collisions, $N_{\rm coll} N^Y_{pp}$.
As in our previous work~\cite{Du:2017qkv} we utilize an entropy-conserving thermal fireball expansion 
(with a lQCD/hadron-resonance-gas equation of state) at each impact parameter and collision energy 
(which determine the total entropy via the observed charged-particle multiplicity).
The initial $Y$ numbers, $N_Y(\tau=0)$, in the rate equation (and the total $b\bar b$ number needed 
for the equilibrium limit, $N_Y^{\rm eq}$) are determined from measured cross sections in $pp$ 
collisions, plus additional ``cold-nuclear matter" (CNM) effects.
Specifically, we employ baseline values for EPS09 nuclear shadowing~\cite{Eskola:2009uj} 
at the LHC of up to 15\% and 30\% in central collisions at mid and forward rapidity, respectively, 
and a nuclear absorption cross section of 3\,mb at RHIC to account for the observed $Y$ suppression 
in $p$-Au collisions. We have checked that upon reducing the CNM effects by a factor of 2, the overall
fit quality worsens, with a thinner 95\,\% confidence level region and a slightly weaker extracted
potential. Without CNM effects essentially no solutions were found within a 95\,\% confidence level.

\begin{table}[!t]
\resizebox{0.5\textwidth}{!}{  
\begin{tabular}{|l|c|c|c|}
\hline
Experiment
& Rapidity & Data ($R_{\rm AA}$) & Reference \\
\hline
193\,GeV U-U & $|y|<1.0$ & 1S, 1S+2S+3S & STAR~\cite{Adamczyk:2016dzv} \\
\hline
200\,GeV Au-Au & $|y|<0.5$ & 1S, 2S+3S, & STAR~\cite{Ye:2017fwv} \\
  &   & 1S+2S+3S & \\
\hline
2.76\,TeV Pb-Pb & $|y|<2.4$ & 1S, 2S & CMS~\cite{Khachatryan:2016xxp} \\
\hline
2.76\,TeV Pb-Pb & $2.5<y<4.0$ & 1S & ALICE~\cite{Abelev:2014nua} \\
\hline
5.02\,TeV Pb-Pb & $|y|<2.4$ & 1S, 2S, 3S & CMS~\cite{Sirunyan:2018nsz} \\
\hline
5.02\,TeV Pb-Pb & $2.5<y<4.0$ & 1S & ALICE~\cite{Acharya:2018mni} \\
\hline
\end{tabular}
}
\caption{Summary of RHIC~\cite{Adamczyk:2016dzv,Ye:2017fwv} and 
LHC~\cite{Khachatryan:2016xxp,Abelev:2014nua,Sirunyan:2018nsz,Acharya:2018mni} data 
utilized in our analysis.}
\label{tab_data}
\end{table}
For each parameter set, ($a,b,c,d$), we evaluate the  chi-squared as
\begin{equation}
\chi^2=
\sum_{i=1}^{N}\left(\frac{R_{\rm AA}^{\rm mod}(a,b,c,d)-R_{\rm AA}^{\rm exp}}{\sigma_{\rm exp}}\right)^2 \ ,
\label{eq_chisquare}
\end{equation}
summed over $N$=53 experimental data points, $R_{\rm AA}^{\rm exp}$ 
(cf.~Tab.~\ref{tab_data}), and pertinent model values, $R_{\rm AA}^{\rm mod}$; 
$\sigma_{\rm exp}$ denotes the quadratically combined 1-$\sigma$ statistical and systematic 
experimental error,
\begin{equation}
\sigma_{\rm exp}=\sqrt{\sigma_{\rm stat}^2+\sigma_{\rm sys}^2} \ .
\label{eq_error}
\end{equation}
Assuming that a given model result represents the true values, and that the data are 
normal-distributed around these, the distribution of $\chi^2$ values for given $\nu$=$N-n$, 
$\chi^2(\nu)$, is universal (and normalized) and can be used to define a confidence level. 
We employ a 95\% confidence level which for $\nu$=53-4=49 implies $\chi^2$ values below 
$\chi^2(49)$=66.3; this corresponds to an $\alpha$-value of 0.05, \ie, the integration 
of the $\chi^2$ distribution above 66.3 yields 0.05, or: if the model is correct, there 
is only a 5\% chance that the $\chi^2$-value is above 66.3.

\begin{table}[!t]
\begin{tabular}{|l|c|c|}
\hline
Parameter
& Range & Meaning \\
\hline
$a$ & 1.0-4.0 & $T$ slope of $m_D$\\
\hline
$b$ & 0.0-2.0 & high-$T$ slope of $m_S$\\
\hline
$c$ & 0.0-8.0 & low-$T$ slope of $m_S$\\
\hline
$d$ & 0.0-0.9 & $c$-to-$b$ transition region\\
\hline
\end{tabular}
\caption{Summary of the $n$=4 fit parameters.}
\label{tab_para}
\end{table}
The $\chi^2$ values are computed over a grid of parameters $(a,b,c,d)$ 
(cf.~Tab.~\ref{tab_para}) which encompasses the minimum $\chi_{\rm min}^2$ representing 
the ``best fit" and the 95\,\% confidence hypersurface defined by the maximal 
$\chi_{\rm max}^2$=66.3. In between the grid points the results are emulated using a 
4-dimensional quadratic interpolation mapped onto the $R_{\rm AA}$ values. 

Open HF phenomenology in URHICs, especially the large elliptic flow observed for low-momentum
$D$-mesons at both RHIC and the LHC, requires a large enhancement of the HQ thermalization rates 
over those obtained from pQCD Born diagrams~\cite{Rapp:2018qla}. 
Therefore, in addition to the baseline pQCD quasifree rate, we evaluate scenarios with a 
$K$ factor of 5 and 10 in our statistical analysis (and explicitly show results for the former).

\section{Potential Extraction}
In Fig.~\ref{fig_raa} we summarize our fit results to $Y$ $R_{\rm AA}$'s for a selection of 
ALICE, CMS and STAR data for $K$=5; the bands agree well with the data. A very similar
fit quality is achieved for $K$=1 and 10, with ``best fit" results of 
$\chi^2_{\rm min}$$\approx$46 for all cases. Because of this ``degeneracy", we do not need to 
treat $K$ as an independent parameter. As in our previous work~\cite{Du:2017qkv}, we encounter 
significant discrepancies with the 2.76\,TeV Pb-Pb forward-rapidity data; when arbitrarily 
excluding them from the fit, the $\chi^2_{\rm min}$ drops from $\sim$46 to $\sim$35.

\begin{figure*}[!t]
\begin{tabular}{c}
\begin{minipage}[b]{0.32\linewidth}
\includegraphics[width=1.12\textwidth]{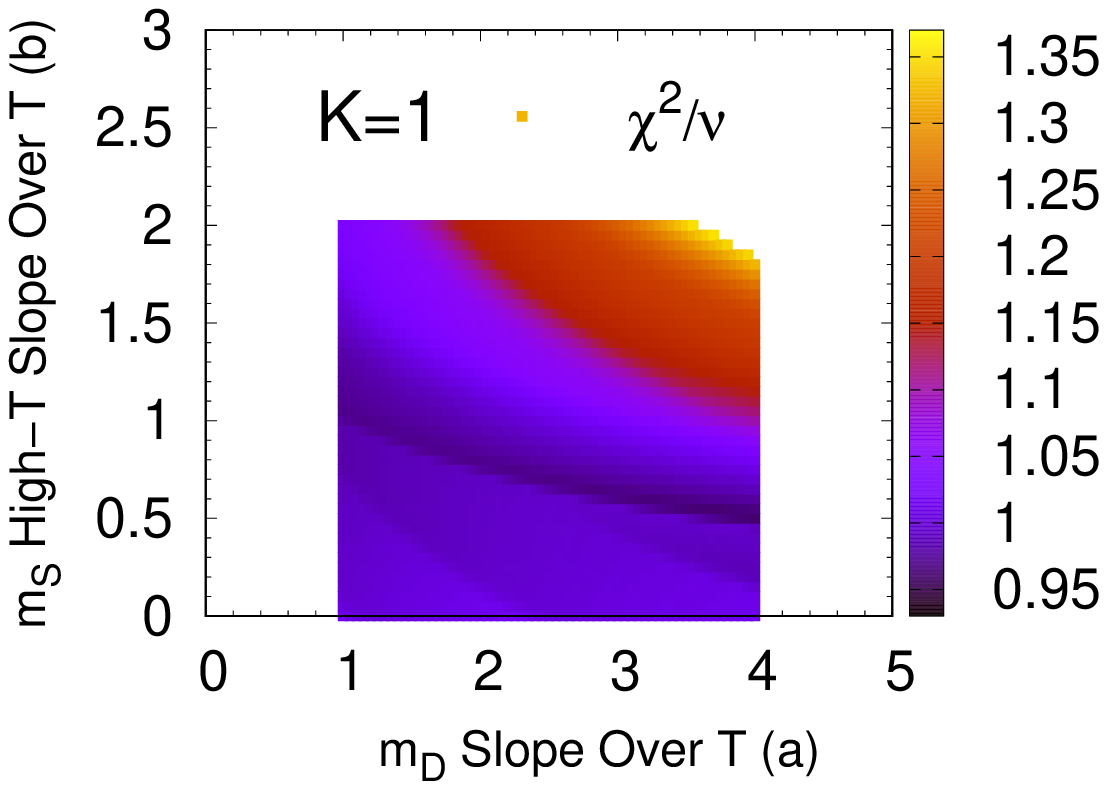}
\end{minipage}
\begin{minipage}[b]{0.32\linewidth}
\includegraphics[width=1.12\textwidth]{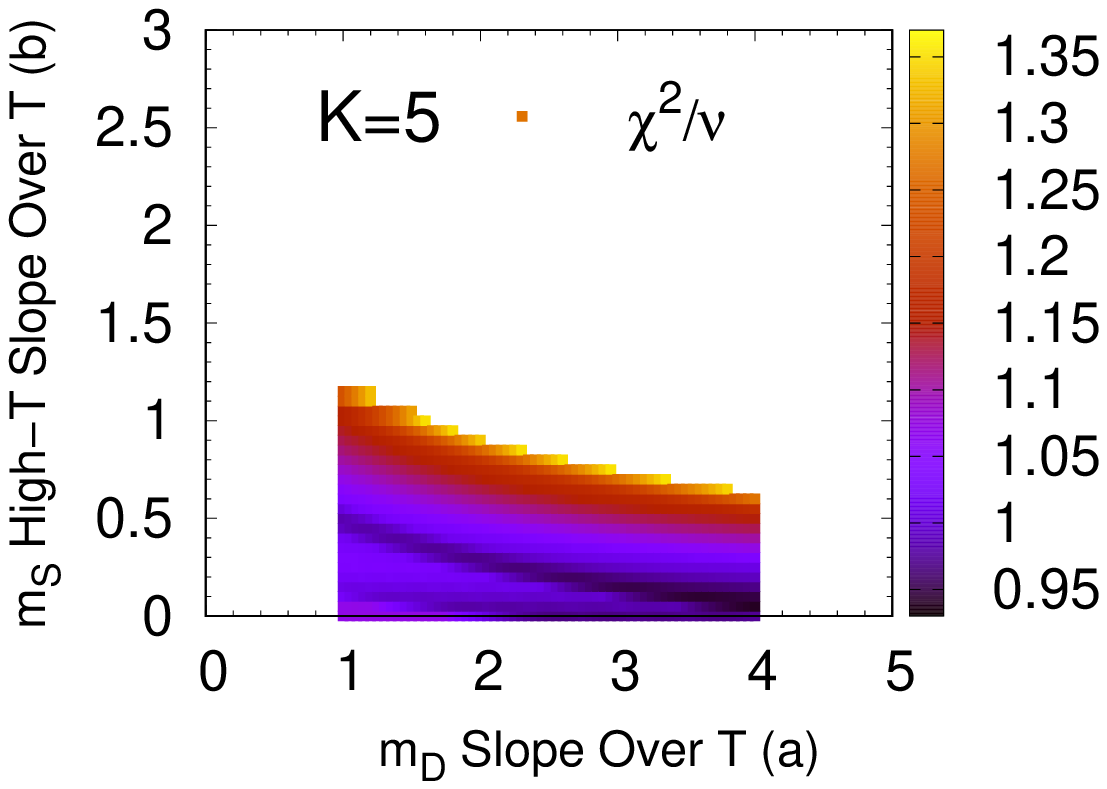}
\end{minipage}
\begin{minipage}[b]{0.32\linewidth}
\includegraphics[width=1.12\textwidth]{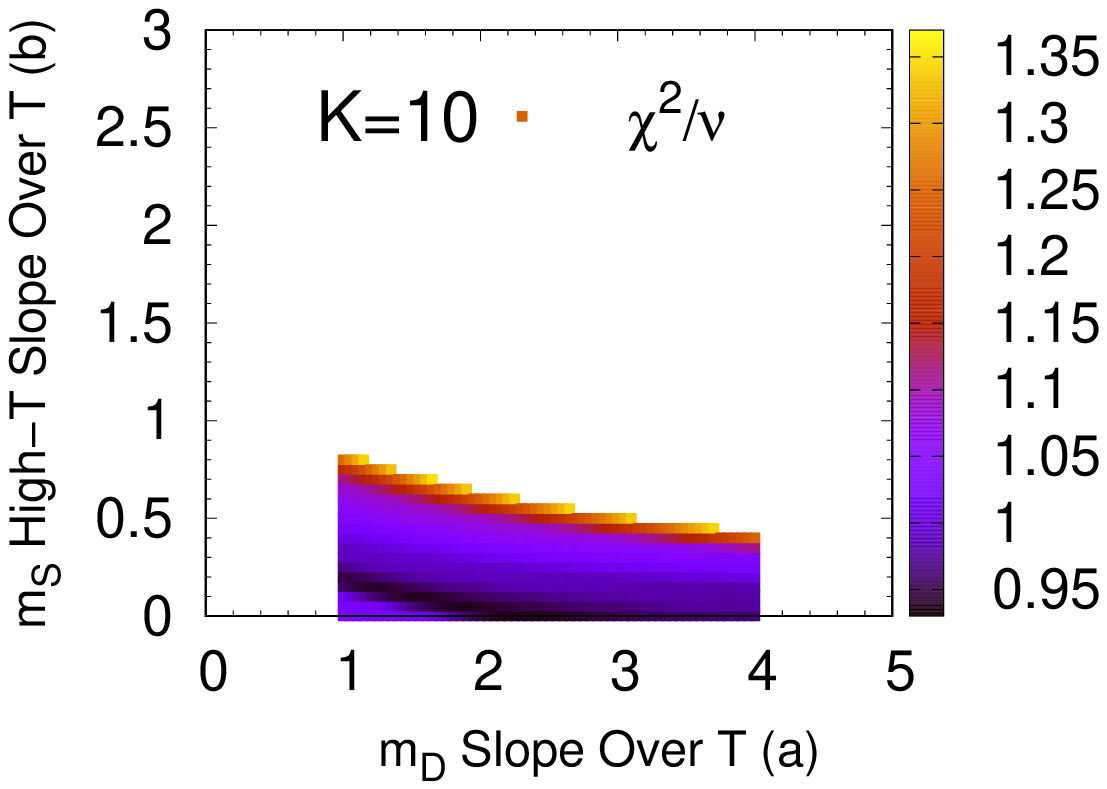}
\end{minipage}
\vspace*{-0.6cm}
\end{tabular}
\caption{Color-coded $\chi^2/\nu$ contours in the $(a,b)$ parameter space (temperature slopes of 
string and Debye screening masses), projected to the minimum values in the associated $(c,d)$ 
space, for $K$=1 (left), 5 (middle)  and 10 (right).}
\label{fig_phase}
\end{figure*}

Inspection of the parameter space in the $(a,b)$ plane (Fig.~\ref{fig_phase}) reveals a 
substantial shrinking of the 95\,\% confidence region of the $T$-dependence of the confining 
force (parameter $b$) as the heavy-light interaction strength ($K$ factor) is increased. At 
moderate temperatures, the increase in the width caused by the $K$ factor is compensated by a 
reduced screening to increase the dissociation energy and lower the final-state phase space. 
On the other hand, the screening of the color-Coulomb potential is not strongly constrained, 
characterized by a large range of values of the temperature slope, $a$, of $m_D$ along 
a valley of $\chi^2/\nu\lsim 1$. This finding highlights the sensitivity of bottomonium observables 
to the confining potential, which is also tightly connected to the strength of the heavy-light 
interaction. Without knowledge of the latter, it is difficult to draw definite conclusions.   

\begin{figure}[!t]
\begin{tabular}{c}
\hspace{-1.0cm}
\begin{minipage}[b]{0.55\linewidth}
\includegraphics[width=1.12\textwidth]{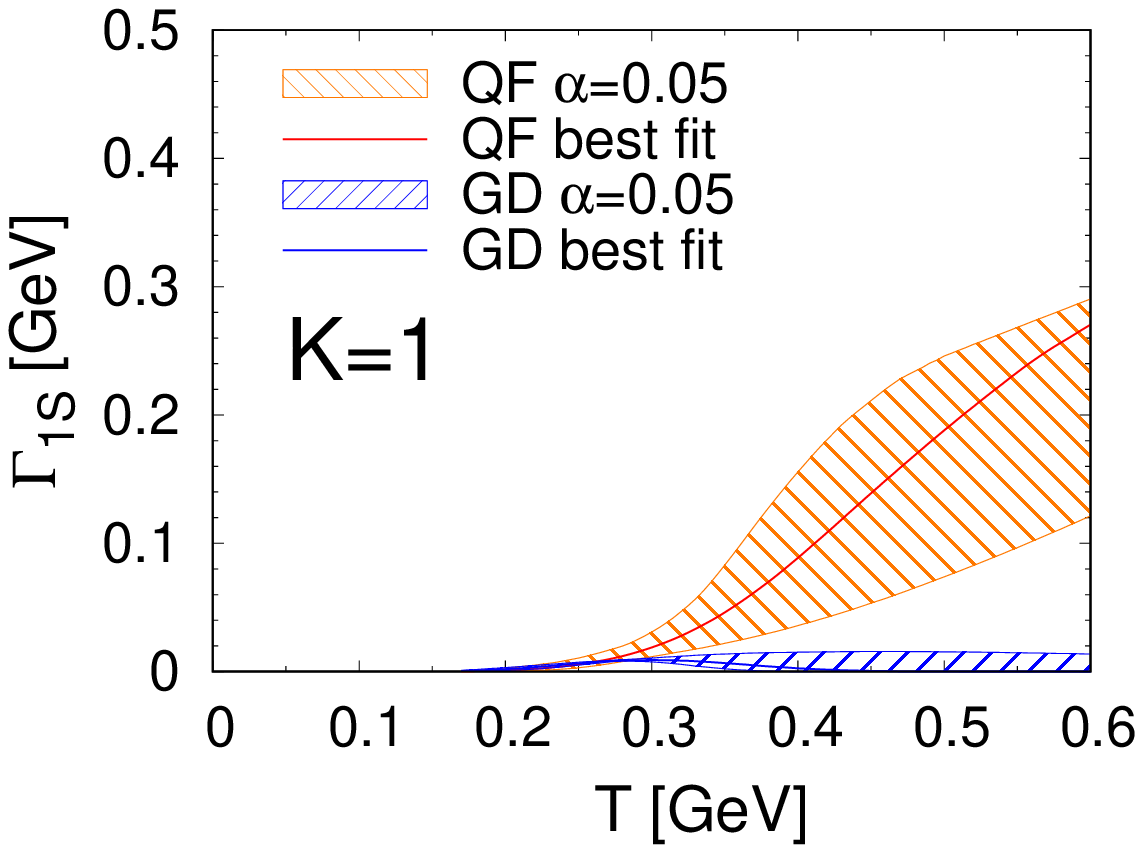}
\end{minipage}
\begin{minipage}[b]{0.55\linewidth}
\includegraphics[width=1.12\textwidth]{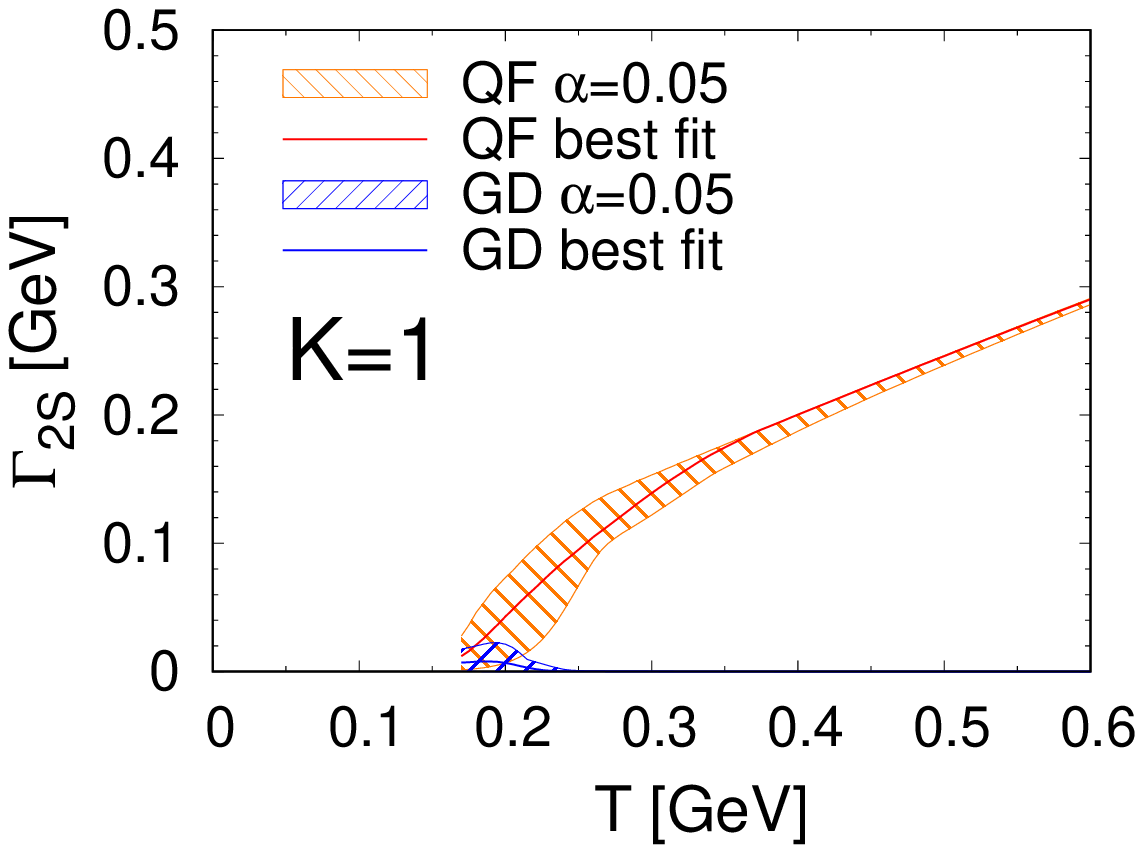}
\end{minipage}\\
\hspace{-1.0cm}
\begin{minipage}[b]{0.55\linewidth}
\includegraphics[width=1.12\textwidth]{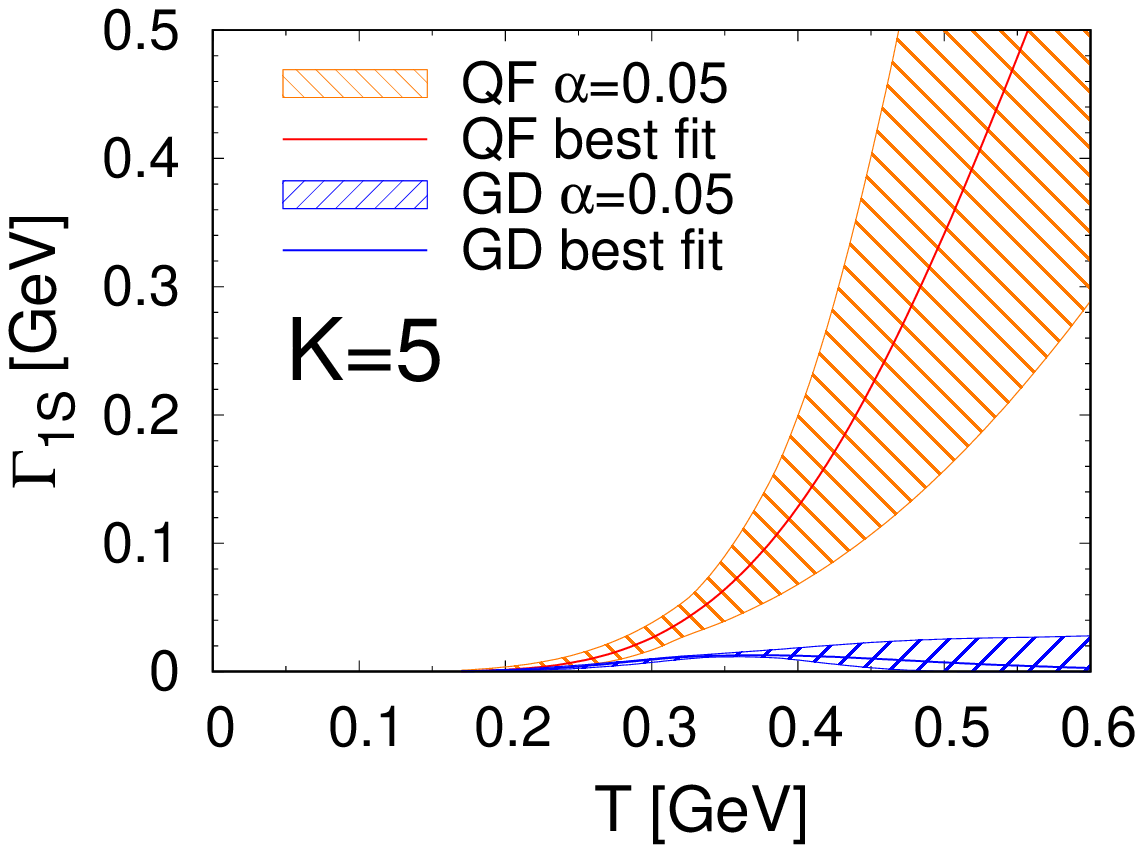}
\end{minipage}
\begin{minipage}[b]{0.55\linewidth}
\includegraphics[width=1.12\textwidth]{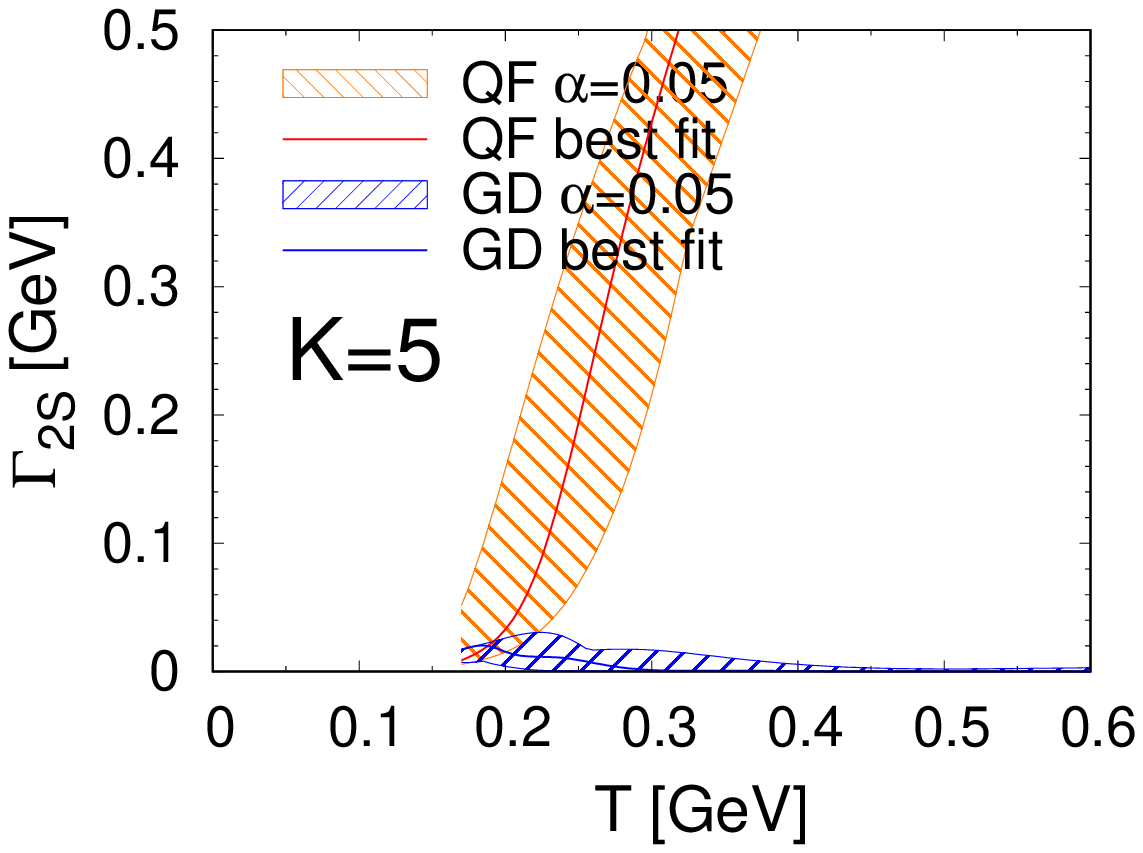}
\end{minipage}
\end{tabular}
\vspace*{-0.6cm}
\caption{95\% confidence bands and best fits (lines) for quasifree (red) and gluo-dissociation (blue) rates for 
$\Upsilon(1S)$ (left panels) and $\Upsilon(2S)$ (right panels) for $K$=1 (upper panels) and $K$=5 (lower panels).}
\label{fig_rate}
\end{figure}

The main transport parameter, the reaction rate, is shown in Fig.~\ref{fig_rate}.
The most relevant temperature region for phenomenology at RHIC and the LHC is $T$$\lsim$400\,MeV since 
the fireball lifetime at higher temperature is (well) below 0.5\,fm/$c$ (based on our previous 
finding~\cite{Du:2017qkv} that the $Y$ $R_{\rm AA}$'s are rather insensitive against variations 
in the initial QGP formation time, which controls the initial temperature; this is in part due
to finite $Y$ formation times). In this temperature range, the resulting $\Upsilon(1S)$
widths are very similar for $K$=1 and $K$=5; they also agree with the microscopic calculations in 
the $T$-matrix approach~\cite{Liu:2017qah}. At higher temperature, the 95\,\% confidence bands become 
broad, but still have overlap until $T$$\simeq$~600\,MeV. The case could be made that this region 
can be probed rather sensitively in a future circular collider in the tens of TeV regime.    
On the other hand, the $\Upsilon(2S)$ rates differ largely beyond $T$$\simeq$300\,MeV (reached 
after roughly 1\,fm/$c$ in central Pb-Pb collisions at the LHC), due to the different $K$ factors
at (near) vanishing dissociation energy. Again, this is somewhat mitigated by its finite formation time, 
but in any case, the $\Upsilon(2S)$ is highly suppressed in semi/central collisions (by 90\% or more 
at the LHC) with a good fraction of the final yield due to regeneration which starts at 
$T$$\lsim$250\,MeV, with then comparable rates for $K$=1 and $K$=5.

\begin{figure}[!t]
\begin{tabular}{c}
\hspace{-0.5cm}
\begin{minipage}[b]{0.52\linewidth}
\includegraphics[width=1.12\textwidth]{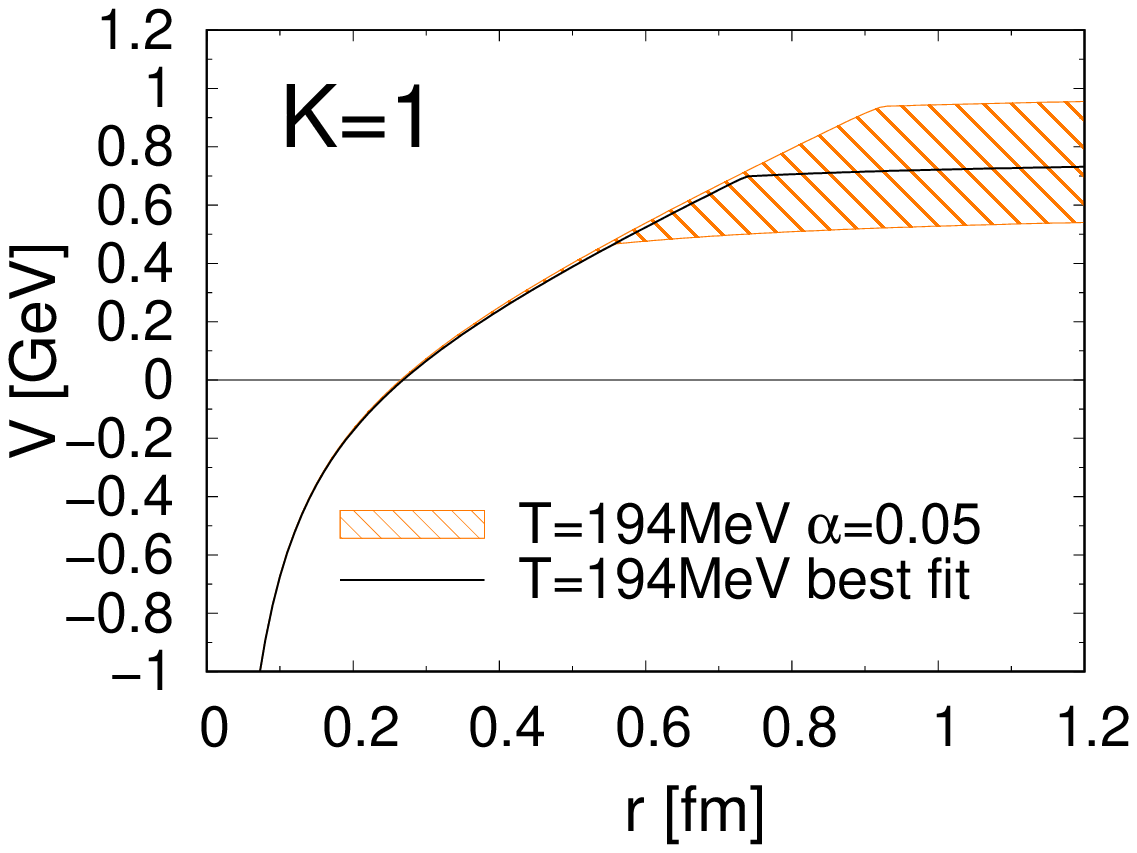}
\end{minipage}
\begin{minipage}[b]{0.52\linewidth}
\includegraphics[width=1.12\textwidth]{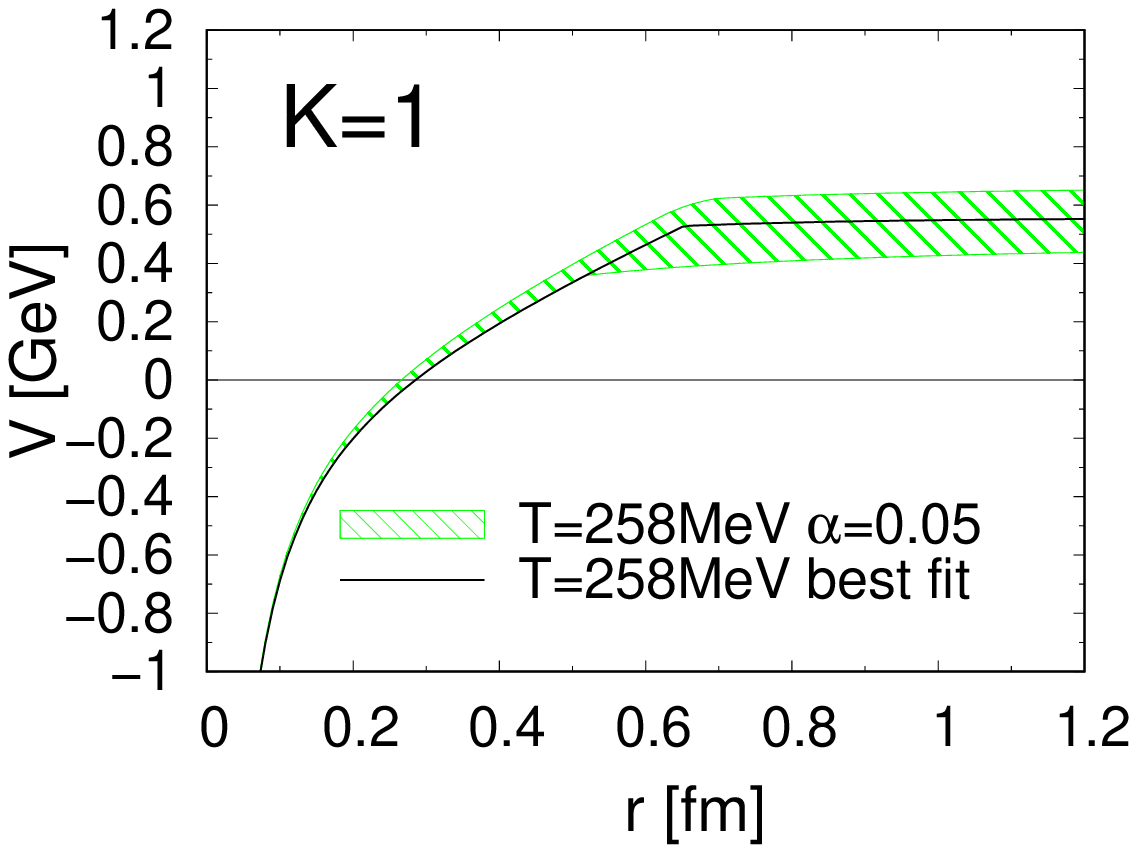}
\end{minipage}\\
\hspace{-0.5cm}
\begin{minipage}[b]{0.52\linewidth}
\includegraphics[width=1.12\textwidth]{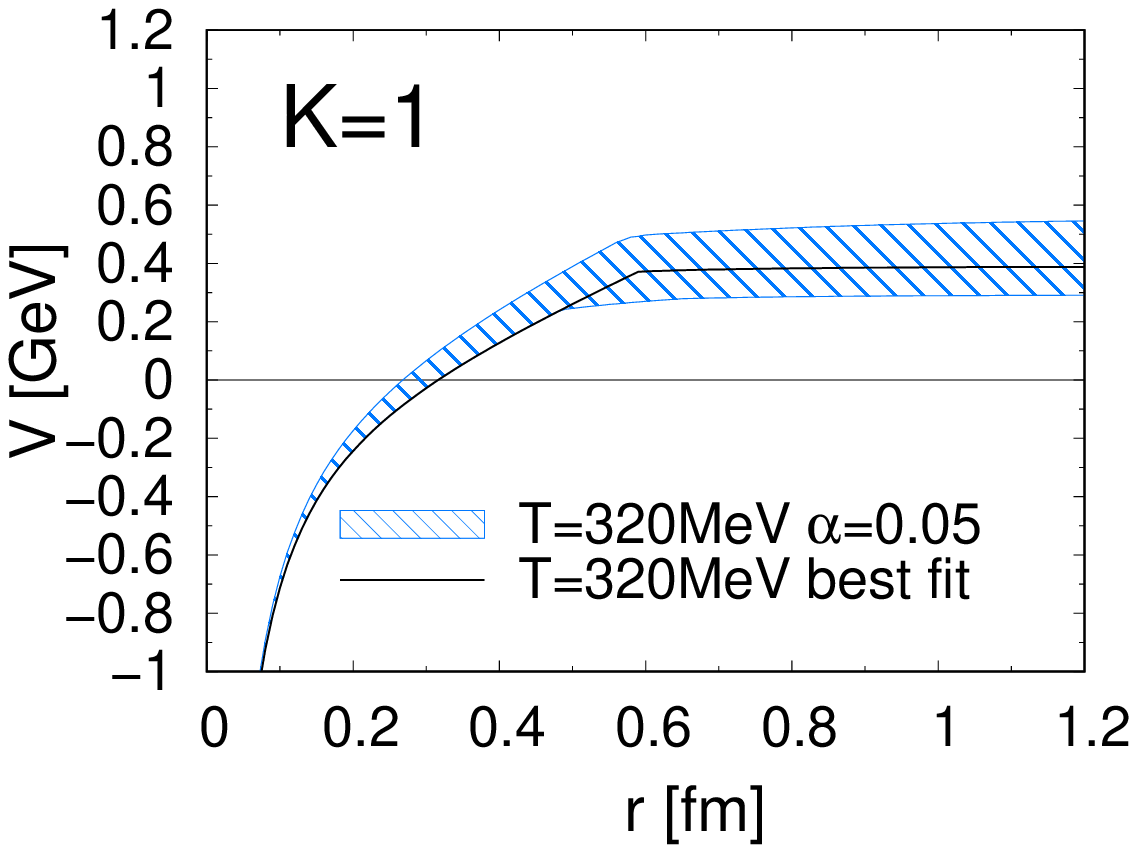}
\end{minipage}
\begin{minipage}[b]{0.52\linewidth}
\includegraphics[width=1.12\textwidth]{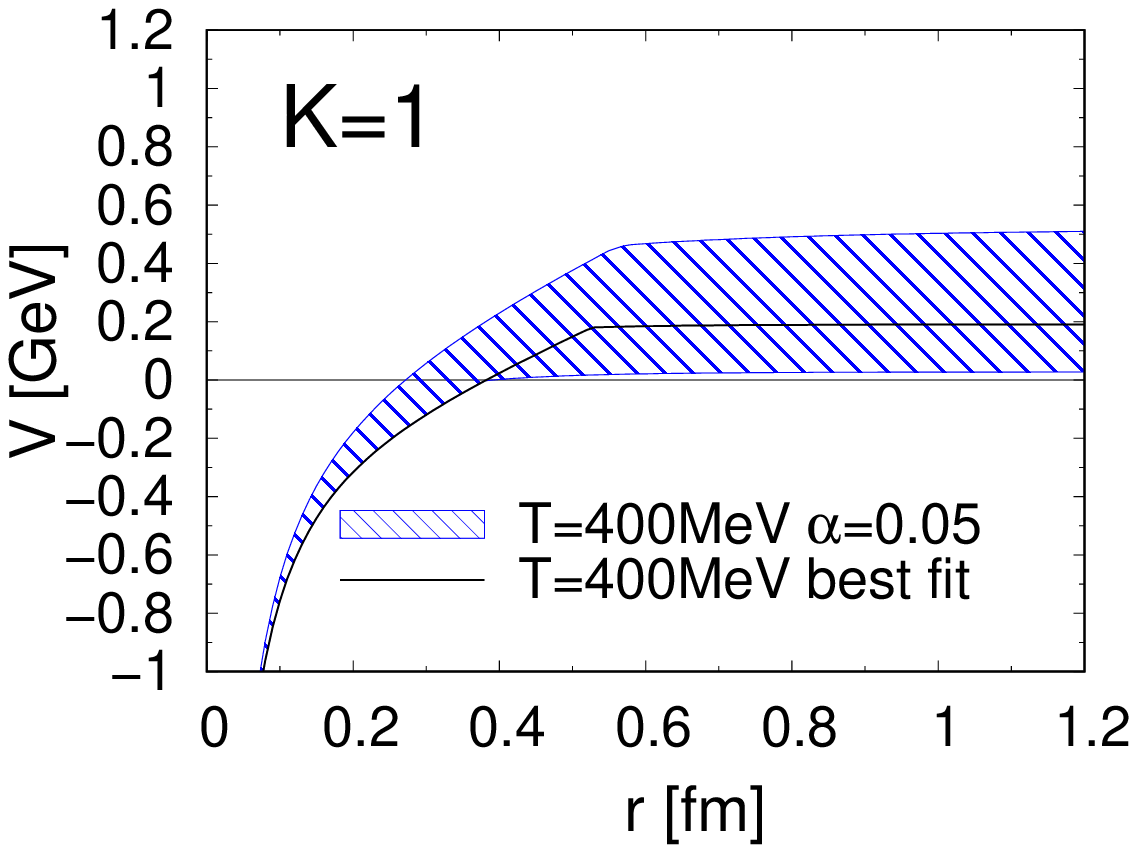}
\end{minipage}\\
\hspace{-0.5cm}
\begin{minipage}[b]{0.52\linewidth}
\includegraphics[width=1.12\textwidth]{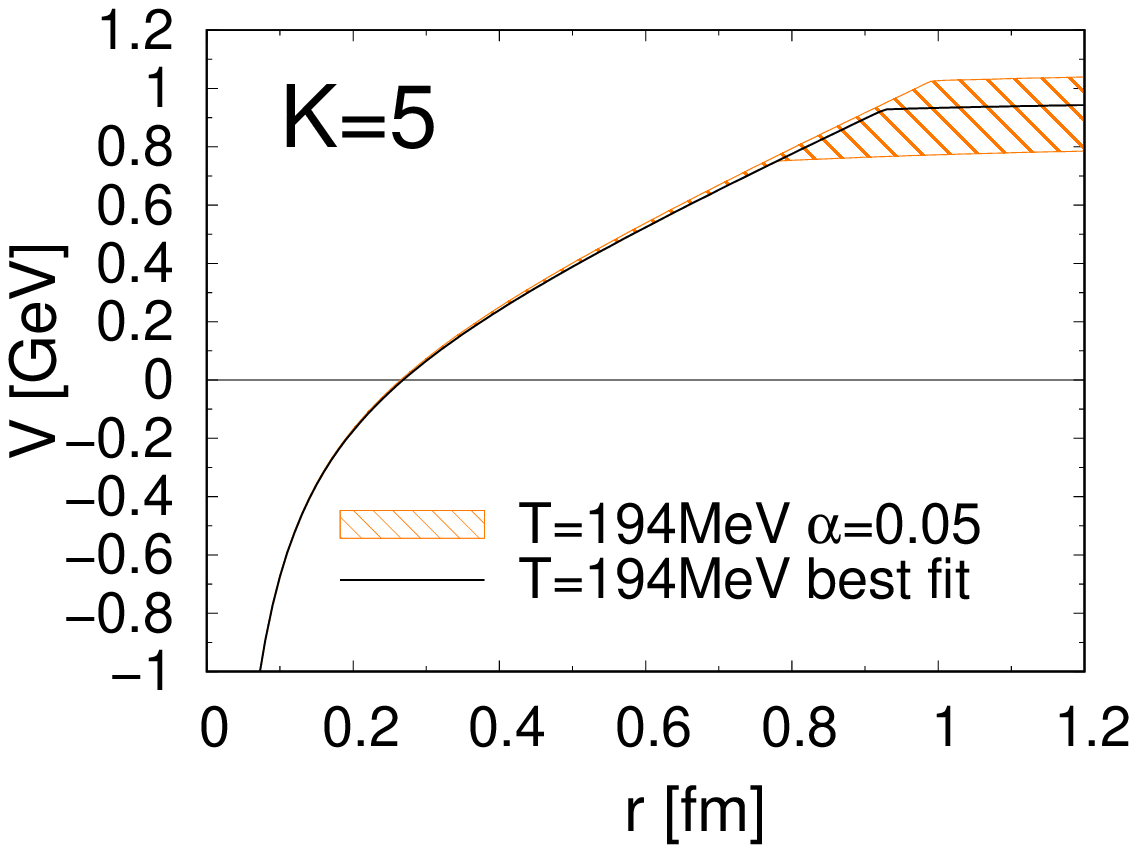}
\end{minipage}
\begin{minipage}[b]{0.52\linewidth}
\includegraphics[width=1.12\textwidth]{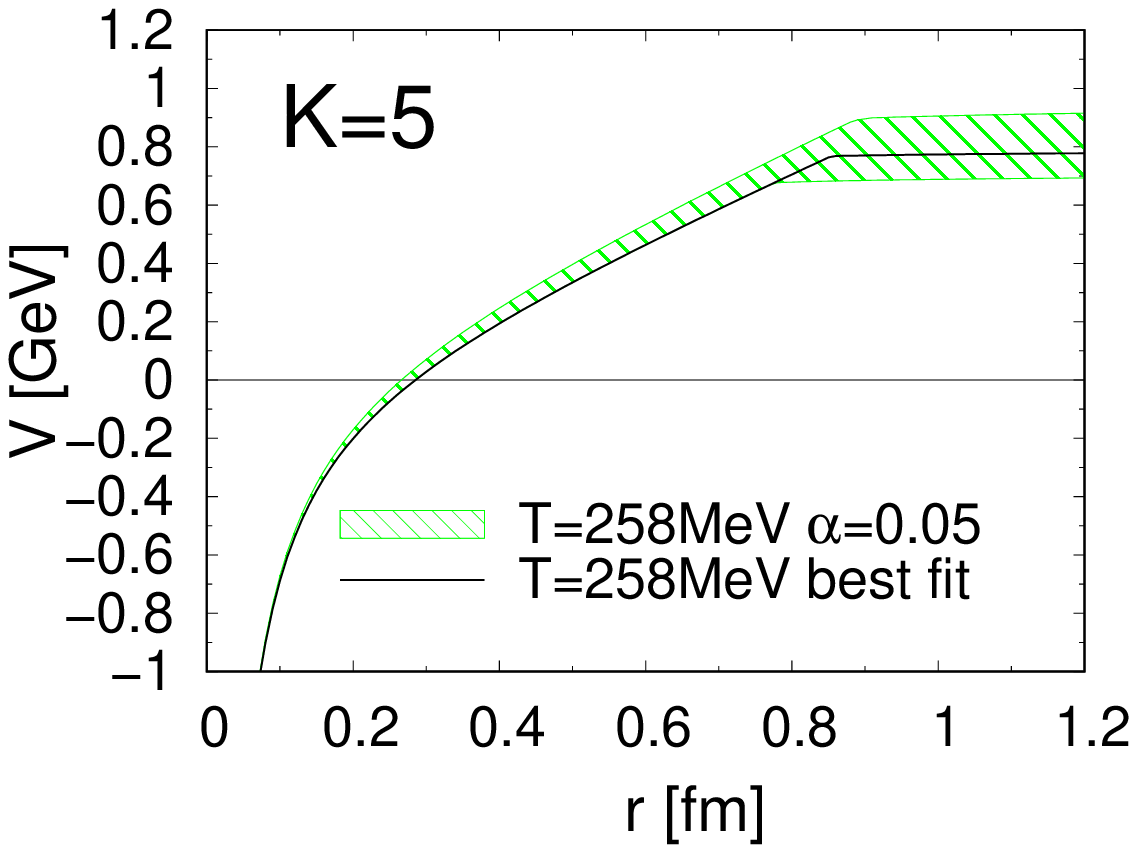}
\end{minipage}\\
\hspace{-0.5cm}
\begin{minipage}[b]{0.52\linewidth}
\includegraphics[width=1.12\textwidth]{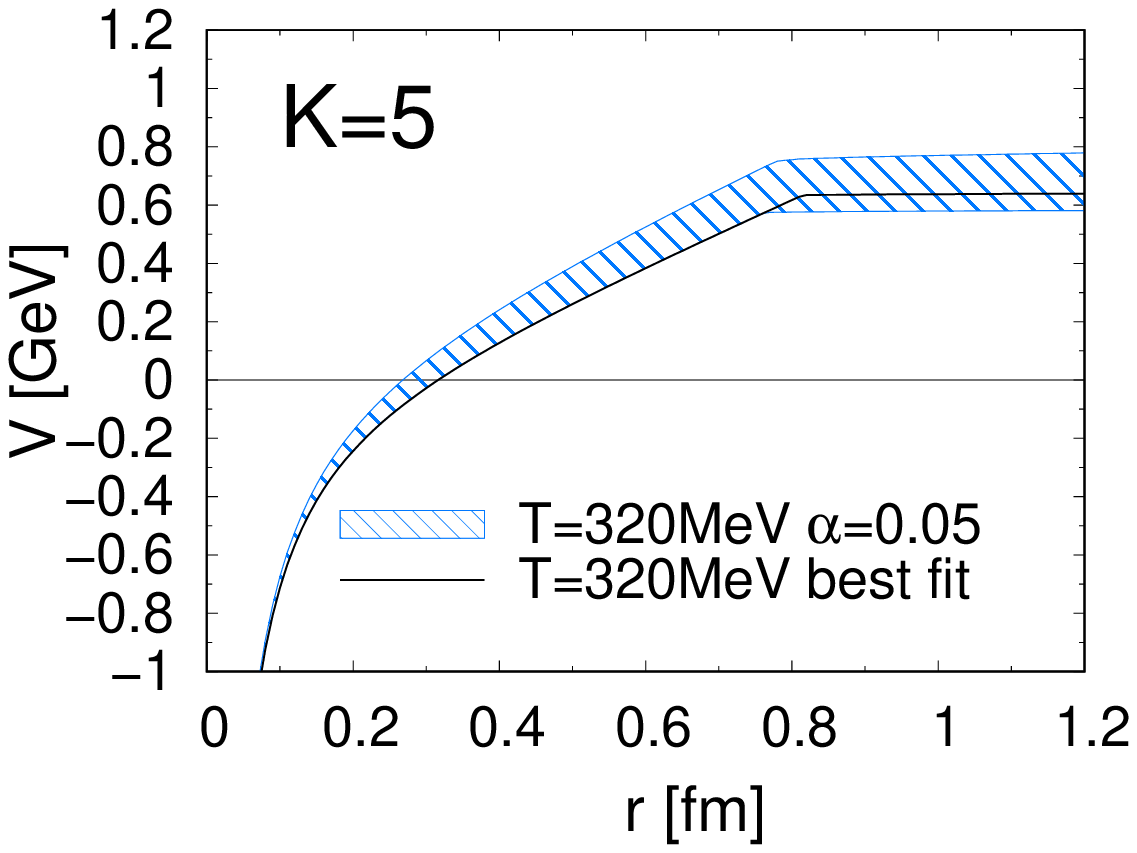}
\end{minipage}
\begin{minipage}[b]{0.52\linewidth}
\includegraphics[width=1.12\textwidth]{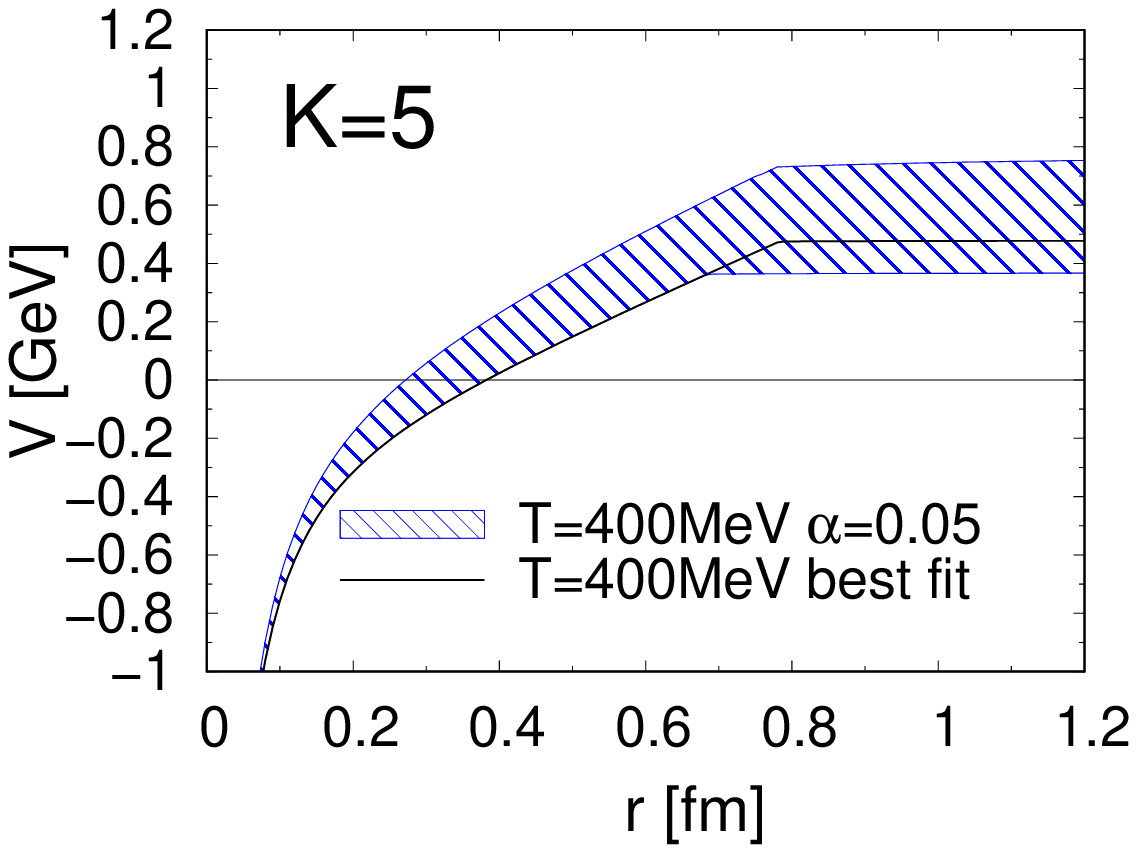}
\end{minipage}
\end{tabular}
\vspace*{-0.2cm}
\caption{95\% confidence level bands for the extracted potential, 
$V(r)=V_{Q\bar Q}(r)-\frac{4}{3}\alpha_sm_D$, and the 
``best fits" (lines) at different temperatures for the $K$=1 (upper 2 rows) and $K$=5 
(lower 2 rows) scenarios.}
\label{fig_pot}
\end{figure}

In Fig.~\ref{fig_pot} we display our main result, \ie, the extracted in-medium HQ potentials at different 
temperatures for $K$=1 and 5. At low $T$, the potentials are close to the vacuum one in both scenarios, 
but for $K$=5 the potential remains substantially stronger at higher temperatures. Since the $K$=1
potential is incompatible with open HF phenomenology~\cite{Liu:2018syc}, the $K$=5 potential should 
be considered a much more realistic solution. Remarkably, the latter closely coincides  with the 
``strong-binding scenario" (with large $E_D$) in the microscopic $T$-matrix calculations of 
Ref.~\cite{Liu:2017qah} which were only constrained by lQCD data (equation of state, quarkonium 
correlators and free energy), not by URHIC phenomenology.

\section{Conclusions}
Utilizing a well-tested quarkonium transport approach, we have conducted a statistical 
analysis to constrain the in-medium heavy-quark potential via bottomonium observables 
in heavy-ion collisions. The potential determines the in-medium $Y$ dissociation energies, 
which in turn govern the reaction rate as the main transport coefficient. 
Guided by theoretical analyses of lQCD data on the HQ free energy, we have employed a 4-parameter 
ansatz to capture essential temperature effects on the color-Coulomb and string force 
components. As an important additional ingredient, we have allowed for a nonperturbative 
enhancement in the bottomonium reaction rates. 
We have then constructed 95\,\% confidence regions of fits to $R_{\rm AA}$ data at RHIC and 
the LHC to extract the in-medium potential for different $K$ factors in the heavy-light 
interaction. The resulting reaction rates essentially coincide in 
the relevant temperature region, dictated by the transport fit to the data, but larger $K$ 
factors lead to significantly stronger extracted potentials in the QGP. The stronger potentials, 
in turn, are required to obtain HQ transport coefficients that are viable for open HF 
phenomenology at RHIC and the LHC.    
Our approach thus highlights the importance of combined analyses of open and hidden HF
probes in a microscopic calculation, and supports earlier independent findings that remnants 
of the confining force above $T_{\rm pc}$ are instrumental for the strong-coupling
features of the QGP.
Several improvements of our work are envisaged. Our previous checks of systematic uncertainties in 
the transport approach (including the bulk medium evolution, $Y$ formation times and the
impact of $b$-quark diffusion), should be revisited, together with explicit calculations of 
nonperturbative effects in the reaction rate~\cite{Liu:2017qah,Liu:2018syc}. This may require the
use of a quantum transport framework as currently being developed from several 
angles~\cite{Akamatsu:2014qsa,Katz:2015qja,Brambilla:2016wgg,Blaizot:2018oev,Yao:2018nmy}, as well as more advanced 
statistical tools to cope with an enlarged parameter space~\cite{Bernhard:2016tnd}. 
Extensions to the charmonium sector, where a rich data set is available, should also be pursued, posing 
additional challenges due to large regeneration contributions and the smaller charm-quark mass.

\section*{Acknowledgments}
This work has been supported by the U.S. National Science Foundation under Grant No. PHY-1614484.

\bibliographystyle{h-elsevier}
\bibliography{refnoda}

\end{document}